\documentclass[aps,pre,
twocolumn,
amsmath,amssymb,amsfonts,
]{revtex4-1}

\usepackage{hyperref}
\usepackage{breakurl}
\usepackage{color}
\usepackage{amssymb}
\usepackage{graphicx}
\usepackage{tabularx}

\begin{document}

\title{Accurate modeling approach for the structural comparison\\
    between monolayer polymer tubes and single-walled nanotubes}

\author{Thomas Vogel}
\email{thomasvogel@physast.uga.edu}
\author{Michael Bachmann}
\homepage{http://www.smsyslab.org/}
\affiliation{Soft Matter Systems Research Group, Institut f\"ur Festk\"orperforschung (IFF-2), 
  Forschungszentrum J\"ulich, D-52425 J\"ulich, Germany}
\affiliation{Center for Simulational Physics, Department of Physics and Astronomy,
  The University of Georgia, Athens, GA, 30605, USA}
\author{Tali Mutat}
\affiliation{Department of Physics, Technion, Israel Institute of Technology,
  Haifa, 32000, Israel}
\author{Joan Adler}
\homepage{http://phycomp.technion.ac.il/}
\affiliation{Department of Physics, Technion, Israel Institute of Technology,
  Haifa, 32000, Israel}

\begin{abstract}
  In a recent computational study, we found highly structured ground
  states for coarse-grained polymers adsorbed to ultrathin nanowires in
  a certain model parameter region. Those tubelike configurations
  show, even at a first glance, exciting morphological similarities to known
  atomistic nanotubes such as single-walled carbon nanotubes. In order to
  explain those similarities in a systematic way, we performed
  additional detailed and extensive simulations of coarse-grained
  polymer models with various parameter settings.
  We show this here and explain why standard geometrical models for atomistic
  nanotubes are not suited to interpret the results of those
  studies. In fact, the general structural behavior of polymer
  nanotubes, as well as specific previous observations, can only
  be explained by applying recently developed polyhedral tube models.
\end{abstract}


\maketitle

\section{Motivation}
\label{motiv}

The considerations of the importance of exact geometric calculations
when dealing with curved nanostructures that we now present
arose from observations we made during a recent
study~\cite{vogel10prl,tv10procathens,tv10proctrond}. There, we
investigated polymers adsorbed at ultrathin nanowires by means of
Monte Carlo simulations applying a common coarse-grained bead--stick
model. For very high adsorption strengths we found, independently of
the effective radius of the nanowire, well-ordered tubelike
monolayer ground-state structures for that system. Those polymer tubes
are formed by aligned helical monomer strands and possess
different chiralities for different tube radii. Similar structural
behavior is known from several tubelike atomic structures with applications in
nanotechnology such as single walled carbon
nanotubes~\cite{dekker98nat,dresselbook,joan1,joan2}.

In order to reveal the morphological similarities between the polymer
monolayer tubes and atomic nanotubes, we first review the common
geometrical view of these structures (Sect.~\ref{classic}{}). We show
in Sect.~\ref{polyhed} that these approximations are not suitable to
explain our findings and how it should be corrected. Finally, we
show in Sect.~\ref{polymer} how exact geometrical calculations of
curved discrete tubes can provide the link between previous
results~\cite{vogel10prl} and results from additional exhaustive and
detailed computational studies of ground states of polymer nanotubes
and real-world atomic nanotubes~\cite{longpaper}.

\section{Review of the traditional approach to nanotube geometry}
\label{classic}

\begin{figure*}
\includegraphics[width=\textwidth,clip]{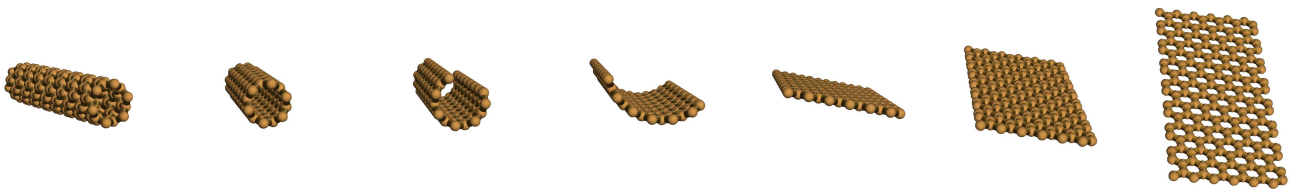}
\caption{\label{fig:movie}%
  Snapshot sequence from an animation illustrating the unzipping of a
  carbon nanotube.}
\end{figure*}

\begin{figure*}
\includegraphics[width=.9\textwidth,clip]{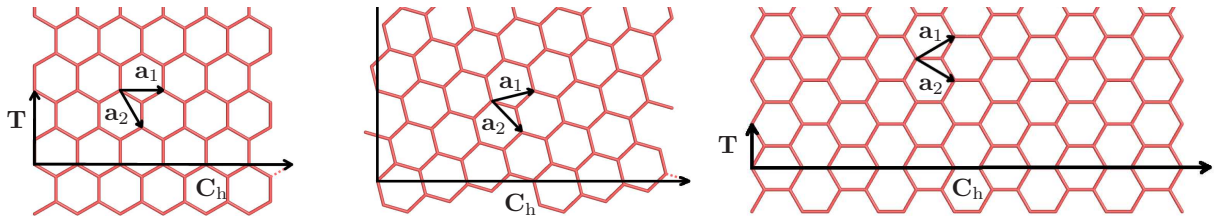}
\caption{\label{fig:planargeom}%
  Definition of the wrapping and the base vectors in three different
  unzipped planar honeycomb structures. Left: (6,0), middle: (6,2),
  right:~(6,6). See text for nomenclature.}
\end{figure*}

A common conception about single-walled carbon nanotubes is that they
are 'built of' rolled up and optionally tilted graphene sheets which
are carbon atoms crystallized in a monolayer honeycomb lattice, as
depicted in Fig.~\ref{fig:movie}. The standard geometrical description
is hence based on this corresponding unzipped planar representation,
which is uniquely defined by a wrapping vector $\mathbf{C}_\mathrm{h}$
pointing from an atomic position to its next periodic copy (see
Fig.~\ref{fig:planargeom}). This vector can be represented as a linear
combination of two base vectors $\mathbf{a}_1$ and $\mathbf{a}_2$ and
two integer numbers $n$ and $m$:
$\mathbf{C}_\mathrm{h}=n\,\mathbf{a}_1+m\,\mathbf{a}_2$. Consequently,
the vector $(n,m)$ is commonly used to classify carbon
nanotubes.~\cite{dresselbook}


In this common picture, the radius $r^{(n,m)}_\mathrm{classical}$ of a
carbon nanotube is calculated by identifying the length of
$\mathbf{C}_\mathrm{h}$ with the perimeter length of the tube:
\begin{equation}
2\,\pi\,r^{(n,m)}_\mathrm{classical}=\left|\mathbf{C}_\mathrm{h}\right|
=a\sqrt{n^2+nm+m^2}\,,
\label{eq:1}
\end{equation}
where $a$ is the edge length in the lattice or the bond length between
carbon atoms, respectively. The wrapping angle
$\theta^{(n,m)}_\mathrm{classical}$ is defined to be the angle between
$\mathbf{C}_\mathrm{h}$ and $\mathbf{a}_1$:
\begin{equation}
\cos\,\theta=\frac{\mathbf{C}_\mathrm{h}\cdot\mathbf{a}_1}{\left|\mathbf{C}_\mathrm{h}\right|\,\left|\mathbf{a}_1\right|}=\frac{2n+m}{2\sqrt{n^2+nm+m^2}}\,.
\label{eq:2}
\end{equation}

Although visualizations sometimes lead to premature assumptions, they
are obviously and doubtlessly quite useful and instructive for the
imagination and interpretation of scientific data, as
Figs.~\ref{fig:movie} and~\ref{fig:planargeom} exhibit.\vadjust{\break}
Figure~\ref{fig:movie} shows snapshots from an animation made with the
\texttt{animate}-package~\cite{animate} for \LaTeX. The input picture
sequence was created using the latest Atomistic Simulation
Visualization software AViz~\cite{aviz,joanproc}.

\section{The polyhedral model and effect of correction terms}
\label{polyhed}

\begin{figure}
\includegraphics[width=.6\columnwidth,clip]{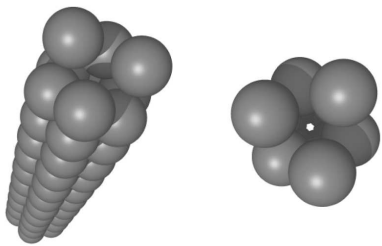}
\caption{\label{fig:30}%
  A triangular nanotube with $(n,m)=(3,0)$, shown from different perspectives. See text for nomenclature.}
\end{figure}

Comparing tubelike ground states of adsorbed polymers and carbon
nanotubes, one first notes that the polymer does not crystallize in a
honeycomb lattice, but rather in a triangular lattice. See
Fig.~\ref{fig:30} showing a triangular (3,0) tube, for example.
However, this does not change much in the above described picture. As
the lattices are dual in a sense, i.e., one can imagine the sites of
the triangular lattice residing in the vacancies of the honeycomb
lattice or centers of the hexagons, one can obviously use the same
notation and calculation as introduced above. The only difference is
that $a$ has to be scaled by a factor of~$\sqrt{3}$.

\begin{figure}[b]
\includegraphics[width=\columnwidth,clip]{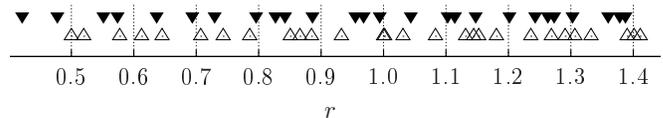}
\caption{\label{fig:vgl1}%
  Possible radii calculated with usual formula (solid, upside-down triangles) and
  radii of defect-free ground-state structures found in simulations (open triangles).}
\end{figure}

However when one compares numerical details, one notes almost
immediately that the above described common approximation does not
provide a suitable model to describe our computational results. To
illustrate this, we present a comparison of radii in
Fig.~\ref{fig:vgl1}. The top row (upside-down solid triangles) shows
the solutions of Eq.~(\ref{eq:1}) for all $n$ and $m$ values
corresponding to $r^{(n,m)}_\mathrm{classical}<1.4$. The bottom row
(upright open triangles) represents radii where we find defect-free
chiral ground states of polymer tubes in our simulations.  Obviously,
there is no apparent link between both sequences. We will illustrate
(assuming that the bond length between monomers or atoms remains
fixed) how the deviations can be explained using two simple examples.
Consider first a triangular (3,0) tube as shown above in
Fig.~\ref{fig:30}, i.e., a tube where three bonds and three monomers
lie in a plane perpendicular to the tube center. With $a=1$, one gets
$r^{(n,m)}_\mathrm{classical}=3/(2\pi)$. However, the three bonds form
a triangle where the sum of the edge lengths equals $3$ and the radius
of the circumcircle reads $r^{(n,m)}_\mathrm{exact}=\sqrt{3}/3$. As a
further example take the (3,3) triangular tube, where
$|\mathbf{C}_\mathrm{h}|=3\sqrt{3}$ and hence
$r^{(n,m)}_\mathrm{classical}=3\sqrt{3}/(2\pi)$. On the other hand the
six corresponding bonds form, projected to a normal plane, a~hexagon
with edge length $\cos(\pi/6)$. The radius of the circumcircle equals
the edge length in a hexagon and therefore
$r^{(n,m)}_\mathrm{exact}=\cos(\pi/6)$. In both examples, obviously
$r^{(n,m)}_\mathrm{classical}\neq r^{(n,m)}_\mathrm{exact}$.

\begin{figure}[b]
  \includegraphics[width=\columnwidth,clip]{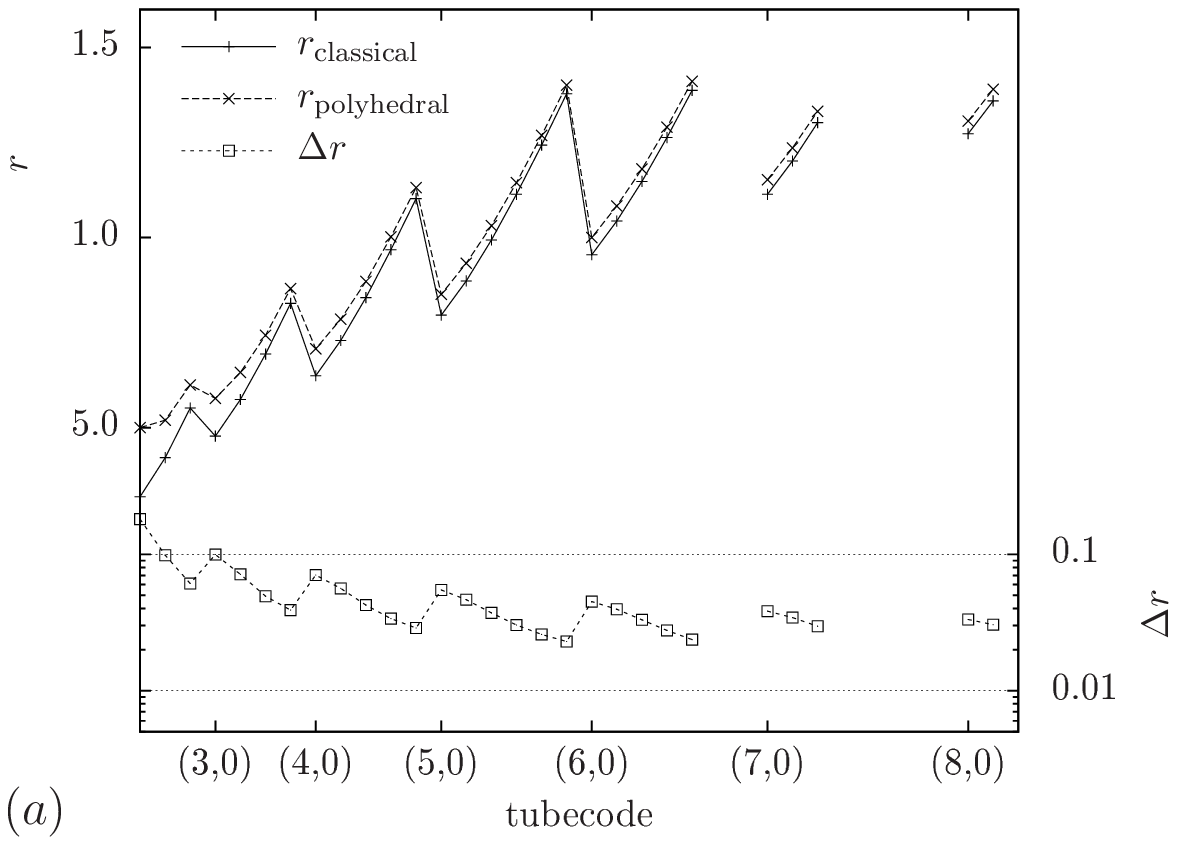}
  \includegraphics[width=\columnwidth,clip]{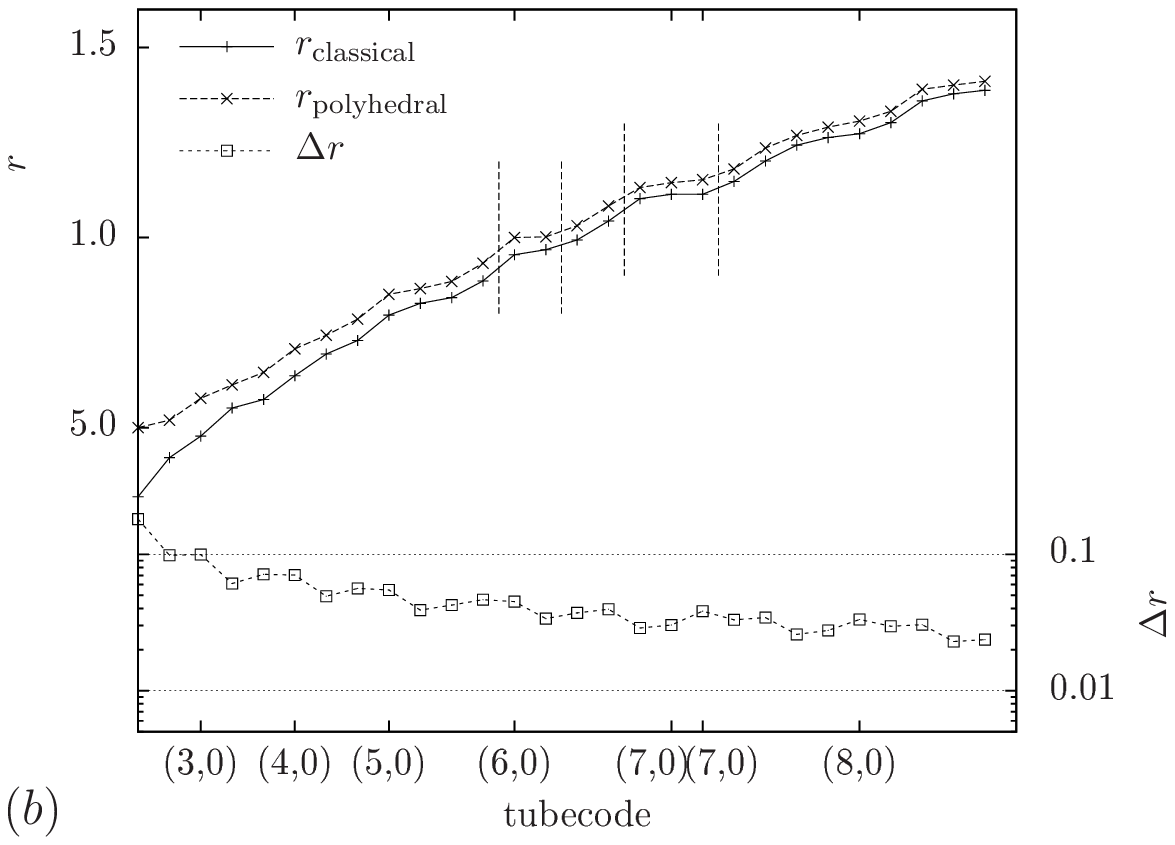}
\caption{\label{fig:vgl2}%
  Comparison of radii of triangular nanotubes calculated in the
    classical and ind the polyhedral model. (a) Data sorted with respect
  to growing~$n$. Data points between $(n,0)$ and $(n+1,0)$
    correspond to $(n,m)$ with $m\in(0\ldots n]$. (b) Data sorted
  with respect to $r$.}
\end{figure}

In fact, when zipping a two-dimensional discrete honeycomb structure,\vspace{5pt}\vadjust{\break}
the geometry changes due to the introduced curvature. Such curved
structures should hence be described by applying a suitable
three-dimensional model rather than an effective two-dimensional one.
For triangular tubes and under the assumption that the edge lengths
remain fixed\footnote{This assumption does not necessarily hold in
  nature, though. See, for example~\cite{budyka05cpl}.}, such a model
is, for example, the polyhedral model introduced for the description of idealized
boron nanotubes~\cite{lee09jpa}. Calculations within this model lead
to transcendental equations for both, the radius of the nanotubes and
the chiral angle, which must be solved iteratively. In
Fig.~\ref{fig:vgl2}, we compare the radii of triangular nanotubes
calculated using the classical and the polyhedral model as well as the
difference for all $n$ and $m$. In Fig.~\ref{fig:vgl2}(a), where the
values of the radii are ordered with respect to $n$ and $m$, those
differences do not seem to be of importance as the qualitative
structure of the curve is the same for both models. But if one sorts
the same data with respect to $r$, as shown in Fig.~\ref{fig:vgl2}(b),\vspace{2pt}\vadjust{\break}
the effect of the deviations becomes apparent. As indicated exemplarily
by the vertical dotted lines, there are intervals where the
differences between the radii of two, in general completely different,
nanotubes is smaller than the difference between the radii for the
same tube calculated with the different models. Hence, one can not
really resolve the link between radius and structure of nanotubes by
using a simple planar ansatz.\footnote{Additionally, the planar ansatz
  can lead to ambiguities as indicated by the the label for the (7,0)
  tube in Fig.~\ref{fig:vgl2}(b). In fact, within that model, the
  radii for the (5,3) tube and the (7,0) tube are exactly the same.}

It should be mentioned that when expanding the relations for radius
and chiral angle~\cite{lee09jpa}, the first term of the expansion is
indeed the same term as one obtains when applying the planar model
(cp. Eqs.~(\ref{eq:1}) and~(\ref{eq:2})). However the higher order terms
are relevant and must not be ignored at least in computational
studies.\footnote{Just for the limiting cases $m=0$ and $n=m$ the higher
  order terms in the calculation of the chiral angle vanish. The
  examples given above for the (3,0) and (3,3) tube are therefore
  correct.}

\section{Structure of monolayer\\ polymer tubes}
\label{polymer}

If we apply the correct polyhedral model for triangular nanotubes, we
find indeed a perfect match between calculated observables and those
found in simulations, in contrast to the situation earlier depicted in
Fig.~\ref{fig:vgl1}. We show the results of both, calculation and
results from simulations, in Table~\ref{tab:1}. In the second column,
calculated radii using the polyhedral model~\cite{lee09jpa} are given, in the
following three columns simulational details can be found and in the last
two columns we list the calculated wrapping angles using a suitable
polyhedral model for carbon nanotubes\footnote{In contrast to the
  planar representation, there is no longer a trivial scaling
  between radii for triangular and honeycomb nanotubes when applying
  the respective polyhedral models. However, the wrapping angles
  are the same.}~\cite{cox07carbon}.

\setlength{\tabcolsep}{1em}
\begin{table*}
\begin{tabular}{lc|c|cc|cc}
\multicolumn{2}{c}{polyhedral} &
\multicolumn{3}{c}{Simulation on Cylinder surface (``2D'')} &
\multicolumn{2}{c}{Corresponding} \\

\multicolumn{2}{c}{polymer tube}&  & \multicolumn{2}{c}{output (Ground State)} &
\multicolumn{2}{c}{carbon nanotube} \\

 & $r^{(n,m)}_\mathrm{exact}$ & $r_\mathrm{input}$ & type & $\theta$ in ${}^\circ$ & $\theta$ in ${}^\circ$ & type \\
\hline
(2,1) &	0.51962 & 0.477\ldots0.532 & 3-helix     & $20.8$\ldots$17.5$  & 18.43& (2,1) \\
(3,0) &	0.57735 & 0.553\ldots0.574 & (3,0)       & $0.0\pm0.5$         &  0.00& (3,0) \\
(2,2) &	0.61237 & 0.585\ldots0.617 & (2,2)       & $31.3$\ldots$29.6$  & 30.00& (2,2) \\
(3,1) &	0.64526 & 0.627\ldots0.670 & 4-helix     & $13.9$\ldots$12.6$  & 13.57& (3,1) \\
(4,0) &	0.70711 & 0.680\ldots0.712 & (4,0)       & $0.0\pm0.5$         &  0.00& (4,0) \\
(3,2) &	0.74313 & 0.723\ldots0.755 & 5-helix     & $23.8$\ldots$22.7$  & 23.33& (3,2) \\
(4,1) &	0.78561 & 0.765\ldots0.808 & 5-helix     & $11.0$\ldots$10.2$  & 10.72& (4,1) \\
(5,0) &	0.85065 & 0.819\ldots0.851 & (5,0)       & $0.0\pm0.6$         &  0.00& (5,0) \\
(3,3) &	0.86603 & 0.861            & (3,3)       & $29.9\pm0.4$        & 30.00& (3,3) \\
(4,2) &	0.88462 & 0.872\ldots0.904 & 6-helix     & $19.2$\ldots$18.3$  & 19.01& (4,2) \\
(5,1) &	0.93259 & 0.914\ldots0.957 & 6-helix     & $9.0$\ldots$8.4$    &  8.84& (5,1) \\
(6,0) &	1.00000 & 0.967\ldots      & (6,0)       & $0.0\pm0.5$         &  0.00& (6,0) \\
(4,3) &	1.00188 & \ldots1.021      & 7-helix     & $26.2\pm0.7$        & 25.26& (4,3) \\
(5,2) &	1.03116 & 1.031\ldots1.052 & 7-helix     & $15.9$\ldots$15.5$  & 16.02& (5,2) \\
(6,1) &	1.08319 & 1.063\ldots1.106 & 7-helix     & $7.6$\ldots$7.2$    &  7.52& (6,1) \\
(4,4) &	1.13152 & 1.106\ldots1.127 & (4,4)       & $30.3$\ldots$30.0$  & 30.00& (4,4) \\
(5,3) &	1.14441 & 1.138            & 8-helix     & $21.7\pm0.4$        & 21.75& (5,3) \\
(7,0) &	1.15238 & 1.148\ldots1.169 & (7,0)       & $0.0\pm0.6$         &  0.00& (7,0) \\
(6,2) &	1.18076 & 1.169\ldots1.201 & 8-helix     & $13.9$\ldots$13.4$  & 13.83& (6,2) \\
(7,1) &	1.23600 & 1.212\ldots1.254 & 8-helix     & $6.7$\ldots$6.4$    &  6.54& (7,1) \\
(5,4) &	1.26887 & 1.244\ldots1.276 & 9-helix     & $26.8$\ldots$26.0$  & 26.32& (5,4) \\
(6,3) &	1.29090 & 1.286            & 9-helix     & $19.0\pm0.5$        & 19.07& (6,3) \\
(8,0) &	1.30656 & 1.297\ldots1.318 & (8,0)       & $0.0\pm0.6$         &  0.00& (8,0) \\
(7,2) &	1.33242 & 1.318\ldots1.361 & 9-helix     & $12.2$\ldots$11.8$  & 12.17& (7,2) \\
(8,1) &	1.39027 & 1.371\ldots1.424 & 9-helix     & $5.8$\ldots$5.6$    &  5.79& (8,1)
\end{tabular}
\caption{\label{tab:1}%
  Comparison of calculated observables for nanotubes using polyhedral models and results from computer simulations of polymers on a~cylinder surface. The rows are ordered with respect to $r^{(n,m)}_\mathrm{exact}$.}
\end{table*}

In order to facilitate the simulations and to obtain more precise
data, we adapted the model with respect to the given problem.  In
contrast to our recent study~\cite{vogel10prl} we introduced flexible
nonelastic bonds between monomers modeled by the FENE potential and
changed the non-bonded Lennard-Jones potential such, that the
equilibrium distances of both interactions match. Furthermore we
initialized the simulation with a configuration where all monomers
have the same predetermined perpendicular distance to the nanowire and
allowed only update moves which not change that distance.  In
practice, we simulated a flexible polymer on cylinder surfaces with
more than one hundred different radii ($r_\mathrm{input}$) and
searched for non-defective ground states, of which we measured the
chiral angle $\theta$. On the one hand, we find the respective
polymers tubes indeed for those radii calculated from the polyhedral
model for triangular nanotubes (cp. columns two and three in
Table~\ref{tab:1}), on the other hand we also measure exactly the
wrapping angles which we calculate for carbon nanotubes (cp. columns five
and six).

Hence, the ground states of monolayer polymers nanotubes forming at
strongly attractive nanowires can be well described by the polyhedral
model for idealized boron nanotubes and the sequence of wrapping angles when
changing the radius of the polymer tube is exactly that calculated for\vadjust{\break}
carbon nanotubes. Our simulations are furthermore very precise, we can
resolve that sequence even for tubes whose radii differ by $<1\%$ (cp.
$r^{(6,0)}_\mathrm{exact}$ and $r^{(4,3)}_\mathrm{exact}$ or
$r^{(5,3)}_\mathrm{exact}$ and $r^{(7,0)}_\mathrm{exact}$), which was
not possible before.

\begin{figure}
\includegraphics[width=.7\columnwidth,clip]{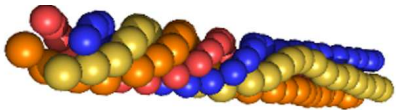}
\caption{\label{fig:bohrer}%
  A ground-state conformation found in a general study of polymers
  adsorbed to ultrathin nanowires. The structure is composed of two
  competing regions with different chiralities.}
\end{figure}

We can also explain specific results from our general study.
Figure~\ref{fig:bohrer} shows a low-energy conformation with tube
radius $0.6<r<0.7$ found for the originally used polymer--wire
model~\cite{vogel10prl}. It shows two competing regions forming a
helix with four strands and a (2,2) structure, respectively. For the
first one we measure a wrapping angle $\theta\approx14^\circ$. Looking
up in Table~\ref{tab:1} we find that that part corresponds to a (3,1)
tube and that this is a direct neighbor of the (2,2) tube with respect
to the possible discrete radii for $(n,m)$ tubes. Indeed, it is
plausible, that there is competition between these structure for radii
that do not match exactly any \smash{$r^{(n,m)}_\mathrm{exact}$}.

\section{Summary}
\label{summary}

In this paper we argued that polyhedral models for nanotubes are
useful for the description of respective structures in computational
studies.  The corrections introduced by those models compared to the
commonly used pictures are in general not negligible (see also a
recent study on the effect of chirality on nanotube
vibrations~\cite{polina}). In particular, the polyhedral model for
boron nanotubes reflects the findings of monolayer polymer nanotube
structures found earlier. The sequence of chiral angles of polymer
nanotubes with different radii is the same as for carbon nanotubes and
provides the link between those structures.

\begin{acknowledgments}
  The authors would like to thank P.~Pine and S.~Srebnik from the
  Technion Haifa for valuable discussions on nanotubes and adsorption
  of polymers at nanotubes. This project is supported by the
  J\"ulich/Aachen/Haifa Umbrella program under Grants No.~SIM6 and
  No.~HPC\_2. Supercomputer time is provided by the Forschungszentrum
  J\"ulich under Projects No.~jiff39 and No.~jiff43.
\end{acknowledgments}

\end{document}